# TOWARDS TRANSPARENT APPLICATION OF MACHINE LEARNING IN VIDEO PROCESSING

L. Murn[1,2], M. Gorriz Blanch[1,2], M. Santamaria[1,3], F. Rivera[1], M. Mrak[1,3]

[1]BBC, Research and Development Department, UK

[2]Dublin City University, Ireland

[3]Queen Mary University of London, Multimedia and Vision Group, UK

**ABSTRACT**

Machine learning techniques for more efficient video compression and video enhancement have been developed thanks to breakthroughs in deep learning. The new techniques, considered as an advanced form of Artificial Intelligence (AI), bring previously unforeseen capabilities. However, they typically come in the form of resource-hungry black-boxes (overly complex with little transparency regarding the inner workings). Their application can therefore be unpredictable and generally unreliable for large-scale use (e.g. in live broadcast). The aim of this work is to understand and optimise learned models in video processing applications so systems that incorporate them can be used in a more trustworthy manner.  In this context, the presented work introduces principles for simplification of learned models targeting improved transparency in implementing machine learning for video production and distribution applications. These principles are demonstrated on video compression examples, showing how bitrate savings and reduced complexity can be achieved by simplifying relevant deep learning models.

**INTRODUCTION**

Machine Learning (ML) has demonstrated superior performance compared to traditional methods when applied to a variety of challenging computer vision and image processing tasks. Methods based on Convolutional Neural Networks (CNNs) have been particularly successful in solving image classification and object detection problems, as well as regression problems including image segmentation, super-resolution and restoration (1).

When applied to visual data, CNNs identify patterns with global receptive fields that serve as powerful image descriptors. The deeper the network, the larger the receptive field, which in turn can lead to the network capturing more complex patterns from the input data. CNNs have set state-of-the-art results in large scale visual recognition challenges (2) and medical image analysis (3). The number of layers of CNN models that have set benchmarks in classification challenges, have been continually increasing, with the *VGG19* model (4) containing 19 layers and *ResNet* (5) containing over 100. These deep architectures act as robust feature extractors and can be used as pre-trained models in related problems. The learned knowledge is applied to a different ML model, raising the accuracy of the related task at hand. For visual content enhancement, applied models such as automatic image colourisation (6), use a pre-trained *VGG19* model for improving the perceptual quality of the outputs, while others like image super-resolution (7), base their architecture on previous

Deep Neural Network (DNN) approaches.

These developments have led to considerable research efforts focused on ways to integrate ML solutions into next generation video coding schemes. Tools based on both CNNs and DNNs with fully connected layers are increasingly being deployed in various newly proposed video compression approaches (8-14).

The increasing demand for video distribution at better qualities and higher resolutions is constantly generating a need for even more efficient video compression. One of the biggest efforts in this field has been related to the development of the next-generation Versatile Video Coding (VVC) standard (15). VVC is a successor to the current state-of-the-art High Efficiency Video Coding (HEVC) standard (16) and aims at providing up to 50% reductions in bitrate for the same objective and subjective video quality. While investigating how to improve such video compression tools by using ML, it has become evident that the main drawback of the application of DNNs is the sheer complexity of their various forms. Moreover, DNN solutions shouldn't be blindly applied to production and distribution applications. Their methods of manipulating input data need to be properly explained and understood, to mitigate potential unexpected outcomes. By making DNNs transparent, it gives an opportunity to build up trust with these methods as we can see what is happening in the network.

To successfully design DNNs for practical applications, we are therefore researching how to address the complexity and transparency issues of DNNs. The approaches presented in this paper utilise specific forms of DNN interpretability which assist in the design of simplified ML video processing tools. As a result, the proposed methods present low complexity learned tools that retain the coding performance of non-interpreted ML techniques for video coding. Furthermore, the methods are transparent and allow instant verification of obtained outputs. With the demonstrated approaches, we have developed and confirmed principles that can serve as guidelines for future proposed ML implementations in video processing.

The paper is structured as follows. In the next section, DNNs are examined, with details of their limitations for deployment in real-time applications. Afterwards, ML interpretability is introduced, with possible benefits for simplifying DNN solutions. A review of various ML proposals for video coding follows, along with a novel approach for efficient learned video processing tools based on ML interpretability. The approach is verified by demonstrating how several existing ML methods can be improved with the proposed enhancements. The conclusion summarizes the developed guidelines for a transparent application of ML in video coding.

**DEEP NEURAL NETWORKS**

DNN architectures typically lead to higher accuracy results for a given task than shallow networks, as seen with large-scale visual recognition challenges (2,4,5). However, researchers have noted that many trained DNNs can be too slow and too large for real-time deployment (17). They advocated a need for fast, compact yet highly accurate models and presented an approach for model compression. The main approach was to use the compact model to approximate the function learned by a slower, larger, but better performing model. While this approach results in only a slight loss in performance, it does not address the transparency issues of learned ML models.

Furthermore, DNN implementations utilizing pre-trained models often have several million parameters. This raises concerns regarding the environmental cost of exponentially scaling these models, while also hampering their adoption rate by having high computational and memory requirements when deployed in applications. Estimates suggest that model training and development likely makes up a major portion of greenhouse gas emissions attributed to many ML researchers, often requiring weeks or months of continuous training on resource-hungry processors (18). The financial costs of these computations can also make it difficult for academics and students to engage in ML research. Such observations lead to recommendations for a concerted effort by industry and academia to promote research into more computationally efficient algorithms. In addition, proposals have been made for adding efficiency of training and of execution as an evaluation criterion for research alongside accuracy and related measures (19). Despite the clear benefits of improving model accuracy, focusing on a single metric, accuracy of the model's outputs, ignores the economic and environmental cost of reaching the reported accuracy.

A study performed on the large visual databases used to train deep classification networks discovered how bias is deeply embedded in most classification approaches for all sorts of images (20). This study prompted many to express concern and distrust towards implementing neural networks within everyday applications. Therefore, it has become very important to explore understanding of neural networks and their learned outputs to rebuild trust. Interpretability, discussed next, is a field which aims to address this.

**MACHINE LEARNING INTERPRETABILITY**

Interpretability is an area of ML research that aims to explain how the results of learned ML algorithms are derived, in a clear and plain manner. Neural network models are commonly used as black boxes, in the sense that it is typically not known how trained algorithms make decisions to arrive at their outputs based on input data. Furthermore, the intricacy of architectures such as CNNs makes them challenging to understand, putting the trustworthiness of their deployment into question. By interpreting neural networks, it is possible to uncover the black box, providing explanations to support a transparent and reliable use of ML. Additionally, this process can also lead to uncovering redundancies in the structure of an analysed model. Understanding the relationships learned by a neural network enables the derivation of streamlined, less complex algorithms that can be applied in systems which require low-complexity solutions and/or do not have enough training data (21, 22).

Previous work showed how interpretability can be used for detecting biases within trained neural network models, in order to be able to avoid deployment of such faulty models. An explanation technique for image classifiers was proposed, demonstrating how ML classifiers can produce undesirable correlations if trained on biased data (23). For example, if a classifier has a task of distinguishing between wolves and huskies, a network can be trained on images where all pictures of wolves have snow in the background, while pictures of huskies do not. Therefore, the trained network will always predict wolf if there is snow in the image and husky otherwise, regardless of the relevant discernible features such as animal colour or pose. Their technique produced saliency maps, which highlighted important features within an image that the network has identified, in this particular case the snow.

Interpretability can also be achieved using post-hoc methods, by analysing the DNN models after training, providing valuable insights into the learned relationships between inputs and outputs (24). Post-hoc analysis captures how much individual features contribute to a prediction. Trained weights of a NN model contain all the parameters needed for performing a series of operations on the input data through each layer of the network. It is very valuable to investigate whether this immense set of parameters can be reduced while retaining the accuracy of the model. This also allows for the understanding of these models, leading to opening up the black box of DNNs and towards their transparent application.

In this paper we focus on techniques that remove certain elements of DNNs that do not affect overall performance, in order to make them more interpretable. The goal is to drastically simplify the application of these learned models while also enabling their transparency. The proposed techniques are verified in the domain of video coding, a research field particularly sensitive to computationally expensive algorithms.

**MACHINE LEARNING IN VIDEO CODING**

Ground-breaking results brought by the usage of ML in visual data tasks have inspired ML applications for video compression. In the last few years, DNN approaches have demonstrated how they can make video compression perform better in terms of bit-rate reduction (8). Traditional tools for predictive coding, filtering or entropy coding can be made more effective when combined with DNN designs (12-14). On the other hand, end-to-end video compression frameworks based on ML outperform the widely used standards, such as H.264 (25). However, such solutions bring coding gains at the cost of substantial increases in complexity and memory consumption. For now, the high complexity of these schemes, especially on the decoder side, limits their potential for implementation within practical applications.

For example, in (12) a proposed attention-based CNN model (aimed at improving the in-loop filtering within VVC) on average achieves bit-rate savings of more than 6% compared to the baseline VVC. When run on a processor, this solution results in over 400 times larger complexity on the decoder side, impairing its possible inclusion in the standard. A few schemes based on highly simplified DNN models have been adopted into the latest VVC drafts, notably Matrix weighted Intra-Prediction (MIP) modes (10) and Low-Frequency Non-Separable Transform (LFNST) (11). In general, although DNNs have shown to bring coding improvements to certain stages of the video compression process, the majority of proposals are still too complex, i.e. too slow for wide-spread adoption.

We have previously proposed complexity reduction techniques for video coding based on learned models (26). In this paper, we concentrate on addressing the efficiency of promising tools which bring considerable coding benefits when implemented within the VVC standard. The improvements in efficiency are achieved by adopting an approach based on ML interpretability. The initial contributions are outlined in (13, 14, 27), and further defined in the following section detailing our approach.

**SIMPLIFYING LEARNED TOOLS APPLIED IN VIDEO CODING**

The predictive coding unit within a video codec removes a video's redundancy in the spatial (intra-prediction) and temporal (inter-prediction) domain. Many ML solutions have been proposed with the goal of improving these tools. These solutions usually utilise fully connected or convolutional layers as its building blocks. Fully connected layers connect all neurons of the previous layer to the next one. This requires an overwhelming number of parameters which can lead to very slow processing, as these parameters are often needed to be accessed thousands of times per second for a single video coding tool during decoding. Convolutional layers are more efficient, analysing features and patterns in visual data by passing a filter over an image, scanning a few pixels at a time.

To achieve impressive performance, NNs require multiple layers to learn abstract relationships within the data. Many of these layers are followed by an activation function that adds a non-linear property to the NN, adding up to a high number of parameters, whether they are CNNs or Fully Connected Networks (FCNs).

By using ML interpretability, we have developed an approach that simplifies these networks, reducing their large number of parameters, but retaining their accuracy. We have applied our approach on three existing ML solutions for video coding:

- Simplification of fully connected layers with application in intra-prediction
- Simplification of convolutional layers with application in inter-prediction

- Branch simplification of complex networks with application in chroma prediction

The methods were verified with the Bjøntegaard delta-rate (BD-rate), a common objective evaluation metric used by video coding researchers (28). Negative BD-rate values represent compression gains, while positive values correspond to compression losses. The BD-rate was computed for a set of video sequences across different resolutions, referred to as classes (29).

**Simplification of fully connected layers with application in intra-prediction**

Intra-prediction reduces spatial redundancies within a frame since adjacent pixels are typically similar. Thus, the content of a given block can be predicted from pixels of its neighbouring blocks.

Conventional intra-prediction modes produce blocks where the predictions are generated from the surrounding samples in a deterministic manner, resulting in even and slow-changing content. ML-generated modes, like MIP modes (10), can produce content that describes more complex textures. An FCN model for intra-prediction (illustrated in Figure 1 (a) was proposed in (27). As the number of parameters of a learned model is high and difficult to interpret, we devised simpler models that are easier to explain. By removing all activation functions and re-training the obtained network architecture, the multi-layer model can be interpreted as a single layer model after training, as seen in Figure 1 (b) and detailed in (30).

Figure 2 shows the depth of both the multi-layer model and the interpreted model. As it's an FCN, the number of parameters for each layer depends on the size of the final predicted square block. The number of reference lines, $D$, is set to 4. Although our network contains four layers, since it's a linear model with no activation functions, it allows for the derivation of the prediction directly from the input. Therefore, it is possible to clearly identify how each reference pixel is used to compute each pixel in a prediction block, as illustrated in Figure 3 for a 4×4 block.

Both approaches were evaluated in VVC Test Model (VTM) 1.0 using all intra configuration, limited to square blocks up to size 16×16. The results show that the simplified model leads to a significantly lower number of parameters, as defined in Table 1. Table 2 shows the coding performance of the proposed approach, achieving up to 1.95% bit-rate savings for the luma channel compared to the constrained VTM, while reducing the complexity of learned FCN intra-prediction by 15% on average (30).

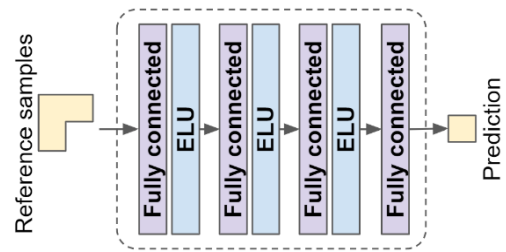

(a) Baseline FCN

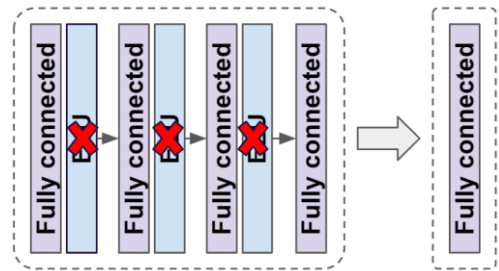

(b) Simplified fully connected model

Figure 1 – FCN simplification process

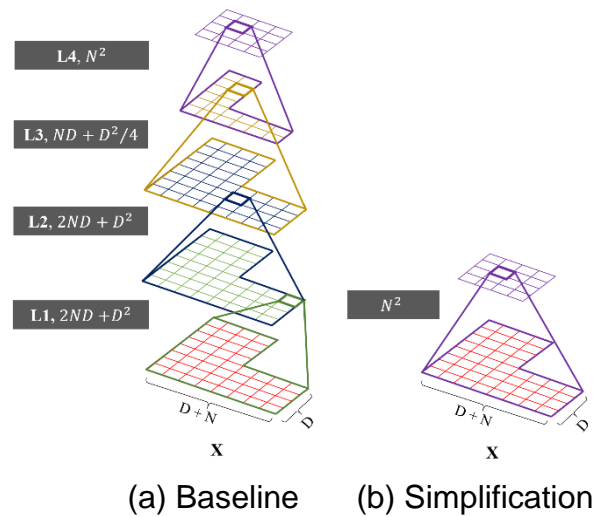

(a) Baseline    (b) Simplification

Figure 2 - Intra-prediction models.

Further simplifications to the approach can be made. The number of reference samples can be reduced as many of them are not used when predicting a pixel (Figure 3).

In general, we have shown that simple ML methods can be derived from complex ones and that the prediction capabilities are similar given that almost identical bit-rate savings were obtained compared to the original model based on FCNs. After demonstrating the effectiveness of interpretability on FCNs, we proceeded to apply our approach on CNN architectures.

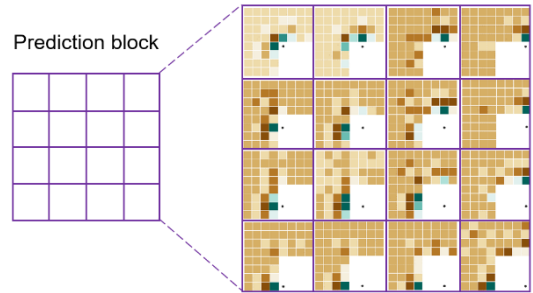

Figure 3 - Contribution of reference pixels to generate a prediction block.

| Network | Parameters |
|---|---|
| Baseline | 92544 |
| Ours | 42751 |

Table 1 - Number of parameters for different FCN structures.

| Class | Luma |
|---|---|
| B | -0.84% |
| C | -1.46% |
| D | -1.71% |
| E | -1.95% |

Table 2 - Coding performance of the simplified FCN model.

**Simplification of convolutional layers with application in inter-prediction**

Inter-prediction obtains a prediction of a block in a current input frame, by utilising a block-matching algorithm to find the pixel-wise closest reference block in a previously encoded frame. Once the best match has been found, the prediction samples from the reference block are subtracted from the original samples. The prediction can be further refined using interpolation filtering to provide a more accurate prediction, known as sub-pixel (fractional) interpolation.

Modern video coding solutions use fixed N-tap filters applied horizontally and vertically to produce fractional samples. However, these fixed filters may not describe the original content well enough or capture the diversity within the video data. Using Super-Resolution CNNs (SRCNNs) to generate new interpolation filters for HEVC was proposed in (13). The method has high complexity requirements, resulting in an almost 50 times higher decoder run-time compared to the HEVC anchor.

In order to see if similar compression benefits can be achieved by using a less complex implementation, we introduced an approach focused on complexity reduction of CNNs, by interpreting the results learned by the networks (31). In the proposed approach, the SRCNN is adapted by removing non-linear activations (Figure 4), which did not impact the learning and compression performance.

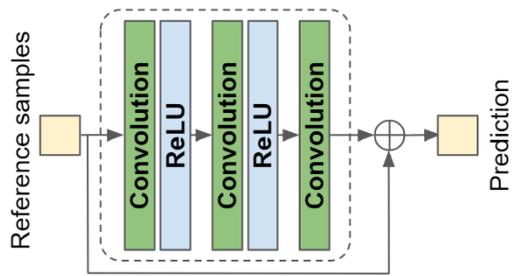

(a) Baseline SRCNN

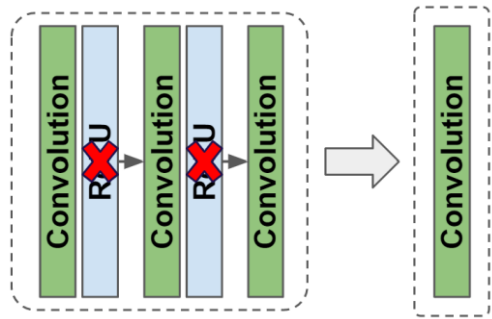

(b) Simplified convolutional model

Figure 4 – CNN simplification process

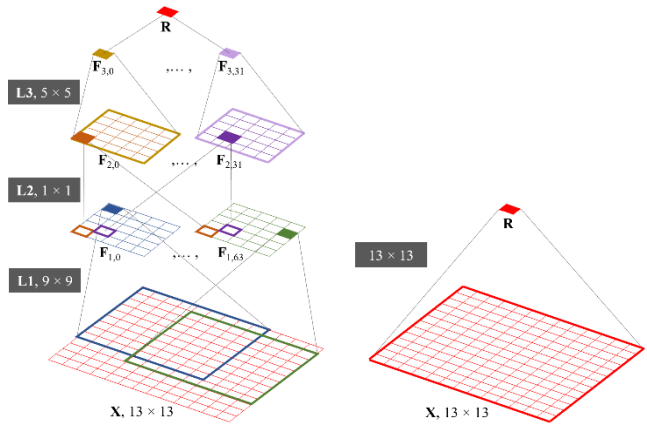

(a) Baseline    (b) Simplification

Figure 5 - Inter-prediction models.

As illustrated in Figure 5, each CNN layer extracts features from the previous one with a receptive field until it arrives at the pixel prediction. As it's a CNN, the size of the filters is consistent regardless of the final predicted block size. The first layer uses 64 9×9 filters, the second 32 1×1 filters and the last one 32 5×5 filters. Due to the obtained network architecture, once a trained model is available we can devise a method to directly compute samples of the resulting image from the input, instead of performing numerous convolutions defined by CNN layers.

It is possible to identify the contribution of each reference pixel for a specific filter, as illustrated in Figure 6. New interpolation filters have been learned, 15 in total to account for all possible quarter sub-pixel positions.

The simplification fully describes how the network behaves and presents a substantial decrease in the number of parameters compared to the original, non-interpreted model, as outlined in Table 3. This technique performs significantly quicker than previous CNN-based efforts when tested within VVC. Experiments have revealed an 82% decrease in the decoder runtime compared to the initial SRCNN approach (31).

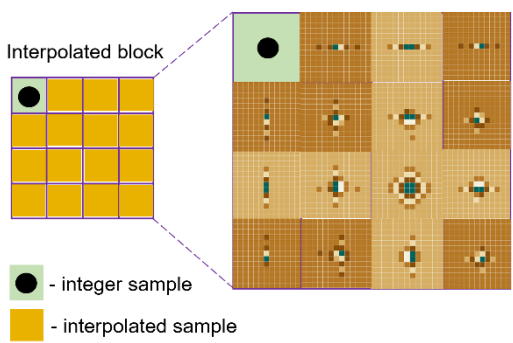

Figure 6 – 15 derived filters, one for each quarter-pixel position

Table 4 summarises test results for a switchable filter implementation (choice between traditional and learned filters) of our approach within VVC constrained conditions. Since the previous method was implemented in HEVC, the encoding configuration was restricted to resemble HEVC conditions for inter-prediction. When compared with the modified VVC, the proposed approach achieves up to 1.1% bit-rate savings in the luma channel for the random access configuration. Additional simplifications of the approach can be made by even further reducing the number of parameters. The derived interpolation filters do not utilise all parameters with equal importance (Figure 6), so some could be removed by quantisation.

After proving that interpretability can lead to less complex NN models while retaining their accuracy, we conducted supplementary research in the area of chroma prediction. We considered more intricate architectures, with a high number of layers, investigating the trade-off between coding performance and memory consumption.

| Network | Parameters |
| --- | --- |
| Baseline | 8129 |
| Ours | 169 |

Table 3 – Number of parameters for different CNN structures

| Class | Luma |
| --- | --- |
| C | -0.65% |
| D | -1.08% |

Table 4 - Coding performance of the simplified CNN model

**Branch simplification of complex networks with application in chroma prediction**

Colour prediction has proven to be effective in achieving better compression rates by exploiting the cross-component redundancies between luma and chroma. The Cross-Component Linear Model (CCLM) was introduced in HEVC (32), as an efficient way of predicting chroma information within an intra-prediction block from already reconstructed luma samples in its close surroundings. Nonetheless, the effectiveness of simple linear predictions can be limiting, and improved performance can be achieved using more sophisticated ML mechanisms. A novel hybrid neural network for chroma intra-prediction was proposed in (14). A CNN for extracting spatial patterns from luma samples was combined with a Fully-Connected Network (FCN) used to extract cross-component correlations between neighbouring luma and chroma samples.

Such techniques improved existing chroma prediction technologies, but increased the system complexity and the amount of operations needed for solving the task. For instance, the aforementioned hybrid architecture required over a hundred thousand operations (including additions and matrix multiplications) for predicting a single block containing 64 pixels. Furthermore, a separate model had to be trained for each possible square block size.

In (33), we introduced a simplified cross-component intra-prediction hybrid model, using interpretability to exploit redundant parameters with the aim of reaching a cost-effective implementation. The proposed architecture introduces a joint model for block sizes of 4, 8 and 16 pixels. It also includes an attention mechanism to control the contribution of each neighbouring reference sample when computing the prediction of each chroma pixel in the current sample location, resulting in more accurate prediction samples. The model complexity is measured by counting the number of parameters. As illustrated in Figure 7, the complexity of utilised convolutional filters can be reduced by removing non-linearities and obtaining a unique filter by combining convolutional layers.

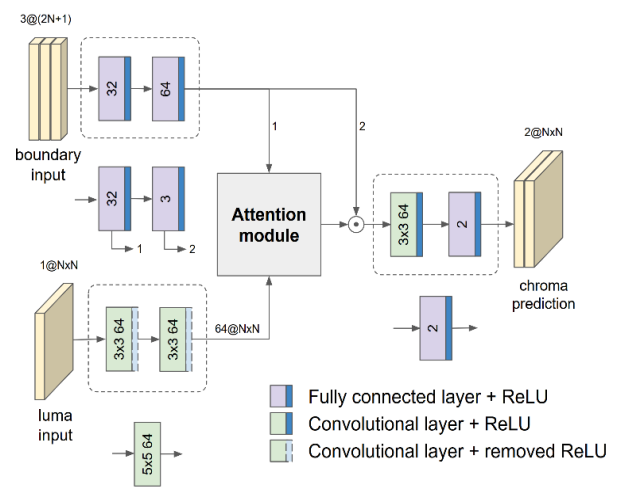

Figure 7 – Chroma intra-prediction model and simplification process.

The derivation process can be seen in Figure 8. Zero padding is applied to convert the input size to $(N+4) \times (N+4)$ pixels. In parallel, the fully-connected branch encodes the colours on the boundary input and transfers them to the unknown block locations. Each boundary input is transformed into a feature vector of 32 dimensions. An autoencoder is used to reduce the branch complexity by squeezing the 32-dimensional feature space into one of 3 dimensions. Autoencoders learn a representation for a set of data, typically in reduced dimensionality, with a trained decoder that recovers the original space. As a result, the coded colours can be obtained using the encoder's part of the autoencoder, hence achieving a reduced complexity. Finally, a fully connected layer is applied to map the output of the attention module to the predicted chroma channels.

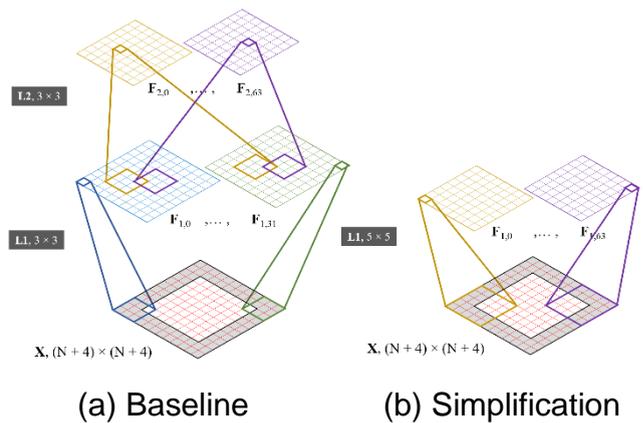

(a) Baseline  (b) Simplification

Figure 8 – Convolutional branch of the chroma intra-prediction model

Table 5 shows a complexity comparison of different architectures for an 8×8 block size. The applied simplifications considerably reduced the system complexity at the price of a minimal performance drop (33). Table 6 summarises coding performance results of chroma channels for a constrained test in VTM 7.0 with square blocks of 4, 8 and 16 pixels. Experiments have revealed 51% decrease of the decoder runtime on average compared to the baseline approach in (33).

| Network | Parameters |
|---|---|
| Baseline | 94116 |
| Ours | 3472 |

Table 5 – Number of parameters for different hybrid network structures.

| Class | Chroma | |
|---|---|---|
| | U | V |
| B | -1.47% | -1.41% |
| C | -1.04% | -1.29% |
| D | -1.13% | -1.64% |
| E | -0.95% | -1.02% |

Table 6 - Coding performance of the simplified hybrid model.

**DISCUSSION**

Through research on the next generation video coding standard, we have validated how existing ML approaches in video coding can be enhanced by using interpretability. We demonstrated three examples that show how certain layers of DNNs can be simplified without a major sacrifice of performance.

In the example of FCN-based intra-prediction, we simplified a network with fully connected layers. By removing the non-linear activation functions, the solution obtained results in a much more compact representation that achieves a sensible trade-off between complexity and coding performance. In the case of CNN-based inter-prediction, we simplified the CNN by also removing activation functions. In addition to previous observations, this approach also enables us to consistently understand how the network processes the input data. In the final example (CNN-based chroma prediction), we established how particular branches of a more complex network can also be simplified without affecting the overall performance.

Although a full simplification may not be attainable, efficiency improvements can be made by understanding the intricate structure of a very deep network.

During this process, we have introduced guidelines towards transparency in applying ML models in video processing. The most important aspect of implementing learned models in production and distribution is understanding how they actually obtain their results. By interpreting these models, we can devise approaches that simplify the methods and also verify their outputs, promoting development of efficient, transparent models with low memory consumption.

**CONCLUSION**

ML algorithms have been increasingly successful in tackling challenging image and video processing tasks. Advancements in technology have allowed for the development of deeper and more complex NN architectures that have led to impressive benchmarks in computer vision. However, their complexity has, at times, also impeded their wide-spread adoption in practical applications. Additionally, ML methods need to be properly explained and understood, especially when applied to production and distribution applications.

Video coding is a research field sensitive to computationally expensive algorithms. Many ML approaches achieve high bitrate savings, but are rarely implemented on resource-limited devices due to their complexity and memory consumption. Our work on interpretable ML examines how the models that underpin these advanced technologies work. We optimise and understand them, in order to leverage that knowledge to improve traditional video coding tools by combining them with DNN designs.

In this paper, we've proven that ML approaches for video coding can be made more efficient by interpreting and thus understanding their learned models. More importantly, the proposed methods can serve as guidelines for subsequent implementations of ML models in real-time production and distribution workflows.

## ACKNOWLEDGEMENTS

The work described in this paper has been conducted within the project JOLT funded by the European Union's Horizon 2020 research and innovation programme under the Marie Skłodowska-Curie grant agreement No 765140, and within an iCASE grant funded by the Engineering and Physical Sciences Research Council of the UK.